\documentclass[global,twocolumn,referee]{svjour}

\usepackage{graphicx}
\journalname{myjournal}
\begin{document}
\title{Generation of long-living entanglement using cold trapped ions with
pair cat states}
\author{M. Abdel-Aty
}                     % Do not remove
%
%\offprints{M. Abdel-Aty: abdelatyquantum@yahoo.co.uk}          % Insert a name or remove this line
%
\institute{Mathematics Department, College of Science, Bahrain
University, 32038, Kingdom of Bahrain }
\date{Received: date / Revised version: date}
\maketitle

\begin{abstract}
{\small With the reliance in the processing of quantum information
on a cold trapped ion, we analyze the entanglement entropy in the
ion-field interaction with pair cat states. We investigate a
long-living entanglement allowing the instantaneous position of the
center-of-mass motion of the ion to be explicitly time dependent. An
analytic solution for the system operators is obtained. We show that
different nonclassical effects arise in the dynamics of the
population inversion, depending on the initial states of the
vibrational motion. We study in detail the entanglement degree and
demonstrate how the input pair cat state is required for initiating
the long living entanglement. This long living entanglement is damp
out with an increase in the number difference $q$. Owing to the
properties of entanglement measures, the results are checked using
another entanglement measure (high order linear entropy).
 }
\end{abstract}

\section{ Introduction}

Controlling quantum systems and the couplings between various
quantum degrees of freedom is one of the major goals in quantum
information processing \cite{pen05}. In the early days of quantum
optics, the dynamics and behavior of typical quantum optical
systems, such as the laser, was dominated by dissipative effects and
uncontrolled fluctuations of system parameters. Systems of single
trapped ions \cite{man05,lan05,jam02} and single atoms strongly
interacting with a high-Q cavity were developed, corresponding to
single quantum systems which can be prepared microscopically and
observed under
conditions close to idealized and fundamental theoretical models \cite{win98}%
. By combining the technologies of ion trapping and cavity quantum
electrodynamics, deterministic coupling of single ions to an optical field
has been achieved \cite{kel05}.

There is currently a diverse array of research interests being united under
the banner of quantum information science, with fundamental research, in
both experimental and theoretical arenas, providing new insights on
different time dependent interaction models \cite{sas02,aty05,aty05a}. Of
great interest to physicists is the renewed vigor with which questions
regarding the basic understanding and interpretation of quantum entanglement
are being pursued. Also, the development of quantum information science and
technology carries the potential of revolutionary impact on many aspects
\cite{zha04}, as evidenced already by the applications in quantum
cryptography, quantum communication, and rudimentary quantum computing.
Entanglement is today considered a fundamental resource in nature when it
comes to quantum computation and information \cite{nie00}, and measures of
entanglement has become a major field of research \cite{skr05}.

Despite much effort and spectacular advances from several groups in
recent years
\cite{ben96,hor98,ved97,ved97b,ved98,fur99,fur01,pho88}, long
surviving entanglement remain elusive in the ion-field based
systems. It is in this light that this paper will focus on
long-living entanglement. The main motivation for the present work
is its relevance to the field of quantum information, which is
attracting broad interest in view of its fundamental nature and its
potentially revolutionary applications. Our main purpose is twofold,
first we wish to demonstrate how a pair cat state affect the
entanglement for the ion-field interaction, and second we wish to
see how the time-dependent amplitude of the irradiating laser field
affect this entanglement. The improved understanding of entanglement
has attracted much attention recently from the quantum information
community which may lead to new insights into the physics of
correlated systems in quantum optics. Therefore there is a great
need to compute these in an efficient way. In particular, we
consider a single trapped ion which can be laser cooled to the
ground state of the trapping potential and discuss the roles played
by the initial state setting and time-dependent amplitude of the
laser field on the entanglement.

The paper is structured as follows: Section 2 is devoted to a brief
description of the pair cat state and show that the quadrature
distribution reflects the entanglement. In section 3, we present the
model of the completely quantized system and obtain an exact
analytical solution of system operators. In section 4, we briefly
discuss those features concerning the atomic inversion, which are
relevant for collapse-revival phenomena. In section 5, we analyze in
detail how the pair cat state affect the general features of von
Neumann entropy (a measure of the ion-field entanglement). This
gives us the opportunity to stress the essential role played by pair
cat states in this context and study the existing of a long living
entanglement. We summarize our results at the end of the paper and
make some conclusions.

\section{Pair cat states}

Superpositions of orthogonal states which exhibits macroscopic
features appear to be of fundamental importance in recent studies of
the foundations of quantum mechanics \cite{buz95,ger97}.
Superposition refers to a quantum system existing simultaneously in
multiple states. This is usually told as the parable of
Schrodinger's cat, shut up in a box with a vial of cyanide that at
any moment might be triggered to release its deadly gas by a
radioactive decay reaction. Resonant microcavities can be used to
study the behavior of mesoscopic superposition coherent states. Some
success has been reported in creating such superposition states
within high Q cavities in the optical domains \cite{mon96}.

Recently, schemes have been presented for the preparation of two-mode
motional states of a trapped ion, such as pair coherent states \cite{gou96},
pair cat states \cite{gou96a}, two-mode SU(1,1) intelligent states \cite%
{ger98}, and SU(2) cat states \cite{zhe00}. These schemes operate in
two-dimensional isotropic traps. In terms of conventional coherent states
parameterized in terms of a complex number $\alpha $ \cite{buz95}, the Schr%
\"{o}dinger cat state has the following form
\begin{equation}
|S_{cat}\rangle \equiv \frac{1}{\sqrt{2+2\cos \phi \exp (-|\alpha |^{2})}}%
\left( |\alpha \rangle +e^{i\phi}|-\alpha \rangle \right) ,
\end{equation}%
where $|\alpha \rangle $ is a coherent state of amplitude $\alpha $, and $%
\phi $ is a real local phase factor. Note that the relative phase $%
\phi $ can be approximately controlled by the displacement operation
for a given cat state with $\alpha >>1$ \cite{jeo02,coc99}.
\begin{figure}[tbp]
%\vspace*{3cm}
%\par
\begin{center}
\includegraphics[width=7cm]{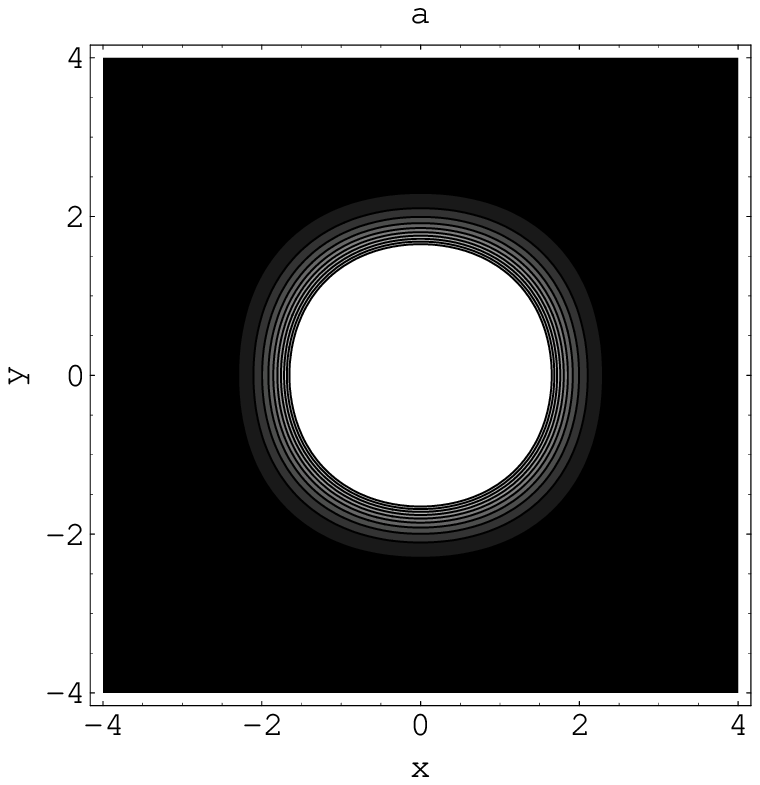} %
\includegraphics[width=7cm]{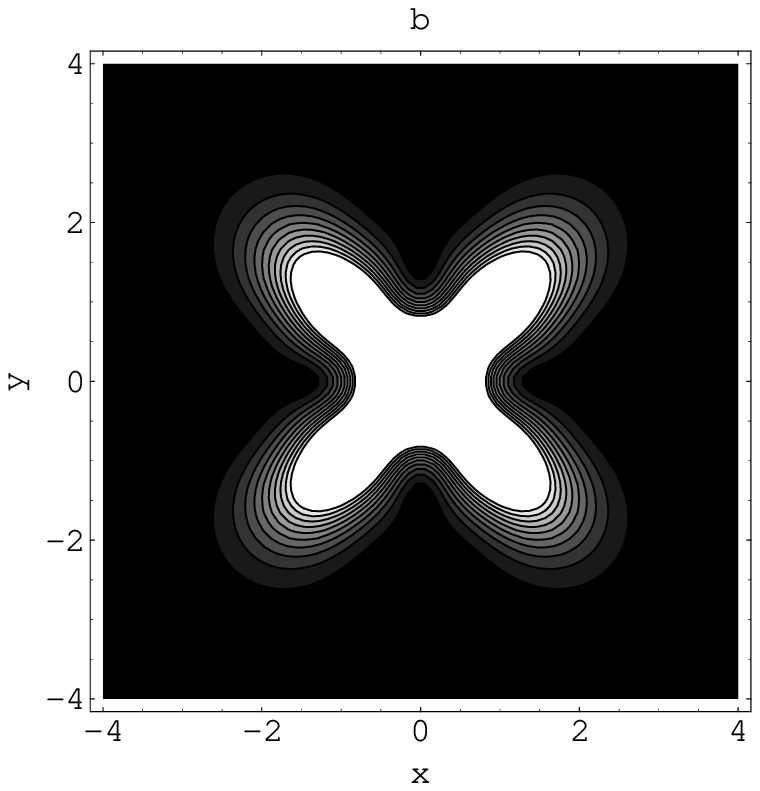} %
\includegraphics[width=7cm]{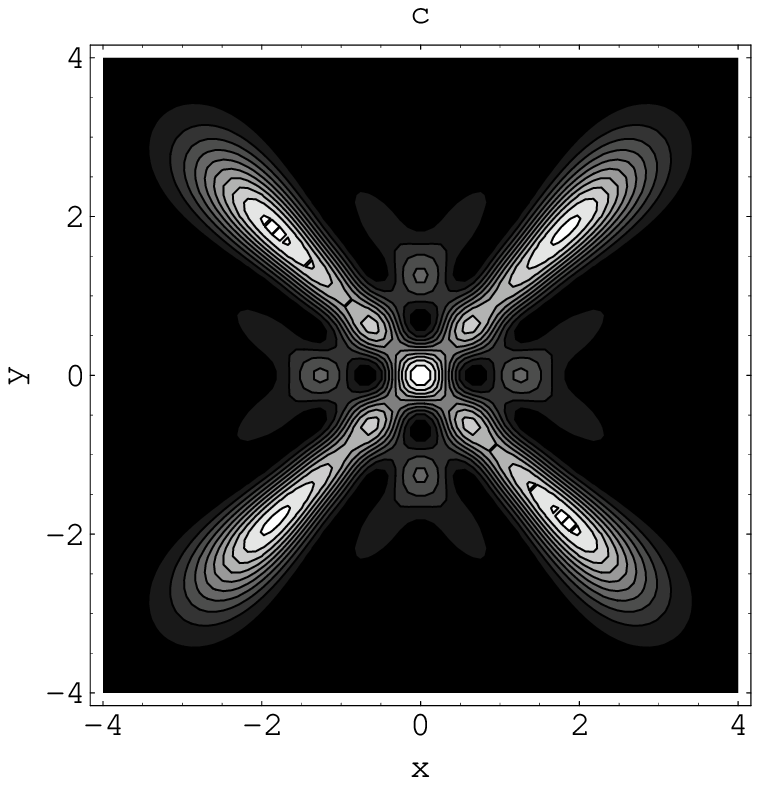}
\end{center}
%\par
% Give a unique label
\caption{Contour plot of the quadrature distribution function
$P(x,y)$ for the pair cat state for $q=0, \phi=\pi/2$ and for
different
values of the parameter $\xi$ where, (a) $\xi=0.1$, (b)$%
\xi=1$ and (c) $\xi=3$.} \label{fig1}
\end{figure}
A special set of coherent states of the Barut-Girardello type known as pair
coherent states has been formulated \cite{kla85}. If $\widehat{a}^{\dagger }(%
\widehat{a})$ and $\widehat{b}^{\dagger }(\widehat{b})$ denote two
independent boson creation (annihilation) operators, then $\widehat{a}%
\widehat{b}(\widehat{a}^{\dagger }\widehat{b}^{\dagger })$ stands for the
pair annihilation (creation) operator for the two modes. The pair coherent
states $|\xi ,q\rangle $ are defined as eigenstates of both the pair
annihilation operator $\widehat{a}\widehat{b}$ and the number difference
operator $\widehat{a}^{\dagger }\widehat{a}-\widehat{b}^{\dagger }\widehat{b}
$ , i.e.,
\begin{equation}
\widehat{a}\widehat{b}|\xi ,q\rangle \equiv \xi |\xi ,q\rangle ,\qquad
\left( \widehat{a}^{\dagger }\widehat{a}-\widehat{b}^{\dagger }\widehat{b}%
\right) |\xi ,q\rangle =q|\xi ,q\rangle
\end{equation}%
where $\xi $ is a complex number and $q$ is the charge parameter,
which is a fixed integer. Furthermore, the pair coherent states can
be expanded as a superposition of the two-mode Fock states,
\begin{equation}
|\xi ,q\rangle \equiv N_{q}\sum\limits_{n=0}^{\infty }\frac{\xi ^{n}}{\sqrt{%
n!(n+q)!}}|n,n+q\rangle ,
\end{equation}%
\begin{figure}[tbp]
%\vspace*{3cm}
%\par
\begin{center}
\includegraphics[width=7cm]{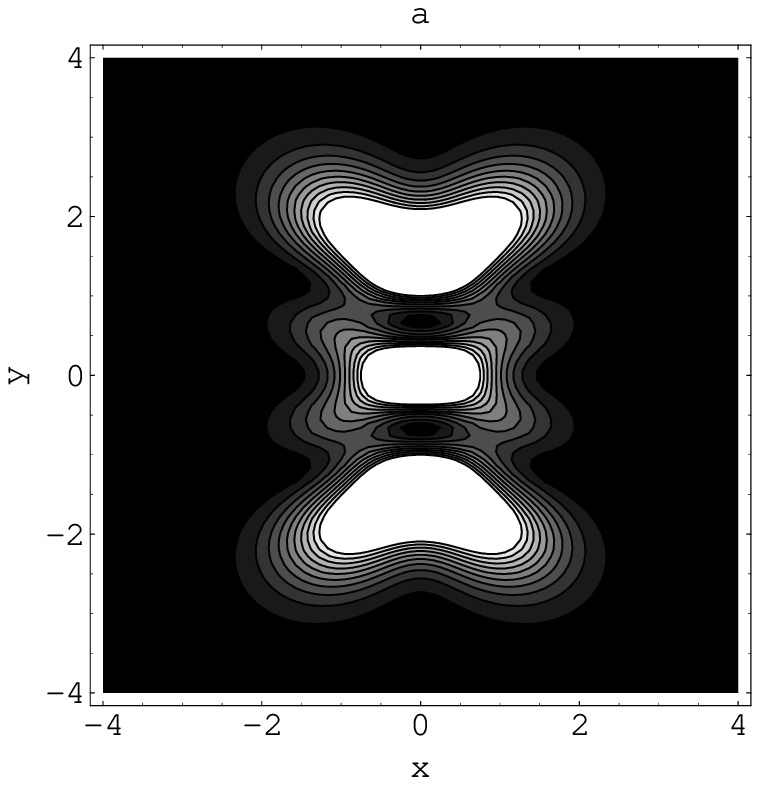} %
\includegraphics[width=7cm]{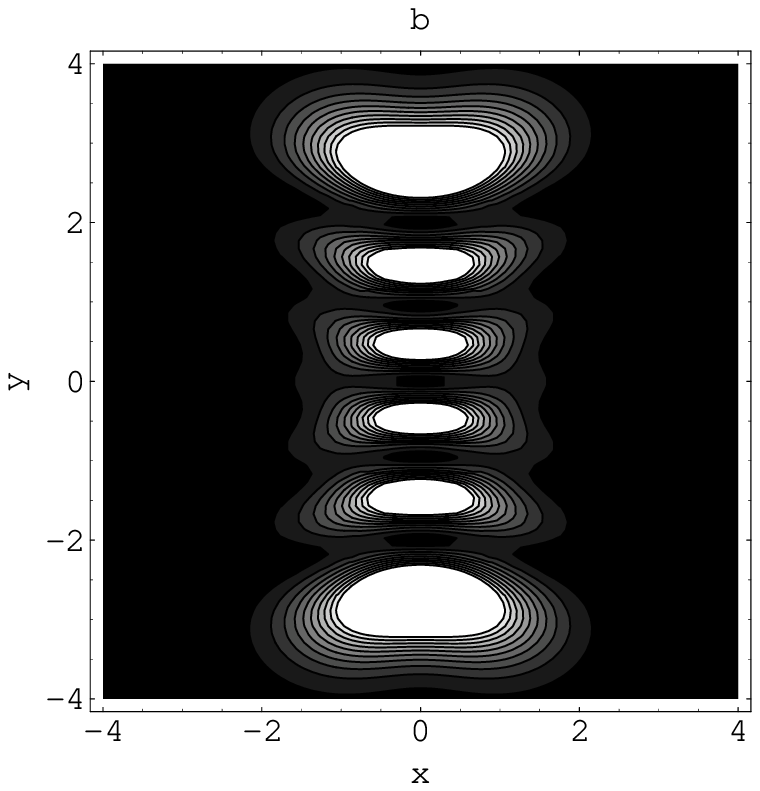} %
\includegraphics[width=7cm]{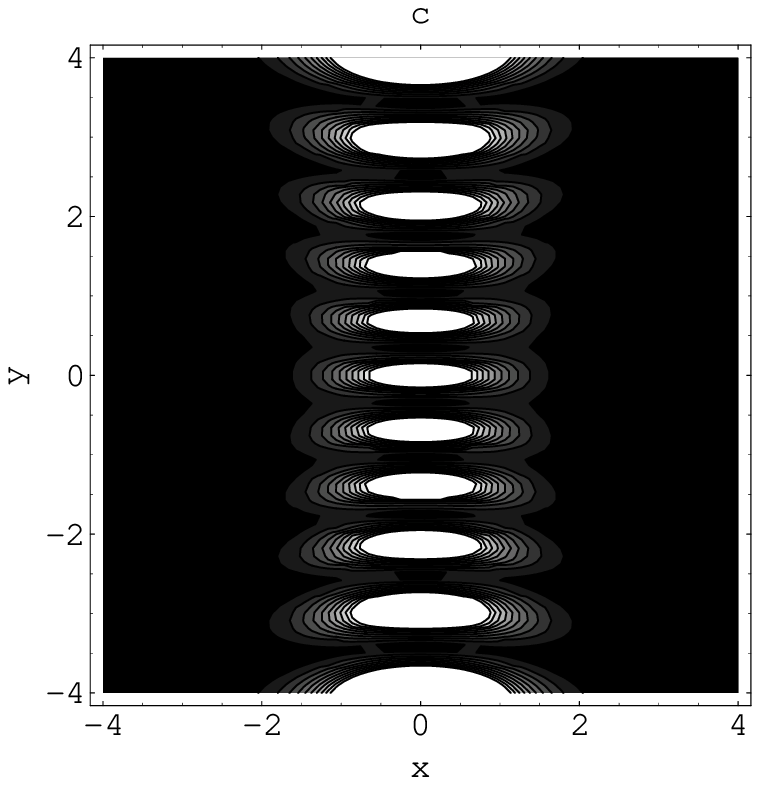}
\end{center}
%\par
% Give a unique label
\caption{Contour plot of the quadrature distribution function
$P(x,y)$ for the pair cat state for $\xi=1, \phi=\pi/2$ and for
different values of the parameter $q$ where, (a) $q=2$, (b)$q=5$ and (c) $%
q=10$.} \label{fig2}
\end{figure}
where $N_{q}$ is the normalization constant and can be written as $N_{q}=%
\left[ |\xi |^{-q}J_{q}(2|\xi |)\right] ^{-\frac{1}{2}},$ where
$J_{q}$ is the modified Bessel function of the first kind of order
$q$. Pair coherent states are regarded as an important type of
correlated two-mode state, which possess prominent nonclassical
properties such as sub-Poissonian statistics, correlation in the
number fluctuations, squeezing, and violations of Cauchy-Schwarz
inequalities \cite{aga88,aga86}. The pair cat states are defined as
a superposition of two pair coherent states separated in phase by
$180^{{\circ }}$ \cite{ger95},
\begin{equation}
|\xi, q, \phi \rangle \equiv N_{\phi}\left(|\xi, q\rangle +\exp
(i\phi )|-\xi, q\rangle \right) ,
\end{equation}%
where the normalization factor $N_{\phi}$ is given by
\begin{equation}
N_{\phi} =\frac{1}{\sqrt{2}}\left( 1+N_{q}^{2}\cos \phi
\sum\limits_{n=0}^{\infty }\frac{(-1)^{n}|\xi |^{2n}}{n!(n+q)!}\right) ^{-%
\frac{1}{2}}.  \label{nphi}
\end{equation}%
It has been shown that these cat states may exhibit stronger nonclassical
effects than the corresponding pair coherent states. The correlated two-mode
cat state $|\xi ,q,\phi \rangle $ is the eigenstate of the operators $%
\widehat{a}^{2}\widehat{b}^{2}$ and $\widehat{a}^{\dagger }\widehat{a}-%
\widehat{b}^{\dagger }\widehat{b}$\ with eigenvalues $\xi ^{2}$ and $q$,
respectively. In order to generate such cat states for the motion in a
two-dimensional anisotropic trap we require two laser beams. Due to the
strong nonclassical nature of the two-mode cat states, the generation of
such states is of interest in testing quantum mechanics.

The coordinate space wave function is given by
\begin{eqnarray}
\langle x,y|S\rangle&=&N_{\phi}\left(\langle x, y|\xi, q\rangle
+\exp
(i\phi )\langle x,y|-\xi ,q\rangle \right)   \nonumber \\
&=&N_{\phi}\sum\limits_{n=0}^{\infty }\left( 1+(-1)^{n}e^{i\phi
}\right) \langle x\left\vert n+q\rangle \langle y|n\right\rangle ,
\end{eqnarray}%
where $\langle x|n\rangle $ is a harmonic oscillator wave function. In this
case, the quadrature distribution is given by
\begin{eqnarray}
P(x,y) &=&|N_{\phi}|^2\left\vert \sum\limits_{n=0}^{\infty
}\frac{\left( 1+(-1)^{n}e^{i\phi }\right) }{\pi
}\frac{2^{-(2n+q)}}{n!(n+q)!}\right.
\nonumber \\
&&\times \left. H_{n+q}(x)H_{n}(y)e^{-0.5(x^{2}+y^{2})}\right\vert ^{2}.
\end{eqnarray}%
For fixed values of the parameter $q$ the periodic nature in the
quadrature distribution function in the different  values of the
parameter $\xi$ are shown in figure 1, where, we plot $P(x,y)$ for
$\xi=0.1,$ 1 and $3$. It
is shown that the regions of the quadrature distribution in the $(x,y)$%
-plane are symmetric with respect to $y=-x$ and with respect to
$y=x$ (figures 1b and 1c). It is clear that the present distribution
reflects the entanglement present in the two-mode pair cat states.
The main curiosity is that, if squeezing is strong enough, the
ring-like quadratic distribution function collapses to a quasi one
dimensional object with a cigar form (see figure 1a).

The origin of such behavior is in the generation of pair cat states
in the microscopic regime, where the contributions of the different
components of the pair cat states are located close to the phase
space origin, competing with each other. When the values of the
parameter $\xi $ is increased, the shape of the quadrature
distribution becomes more pronounced by involving multi-peak
structure (symmetric peaks). This is in a good agreement with the
general behavior of the quadrature distribution of the pair coherent
state \cite{aga05}. It is interesting mentioning here that, the
two-mode pair cat states exhibit strong nonclassical features due to
the correlation between the two modes.
\begin{figure}[tbp]
%\vspace*{3cm}
\par
\begin{center}
\includegraphics[width=7cm]{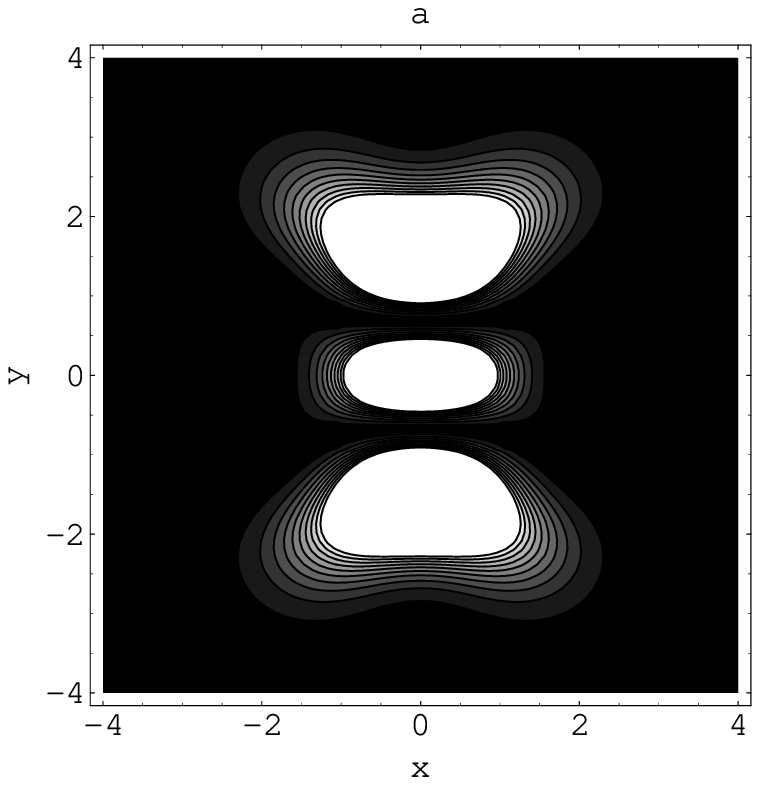} %
\includegraphics[width=7cm]{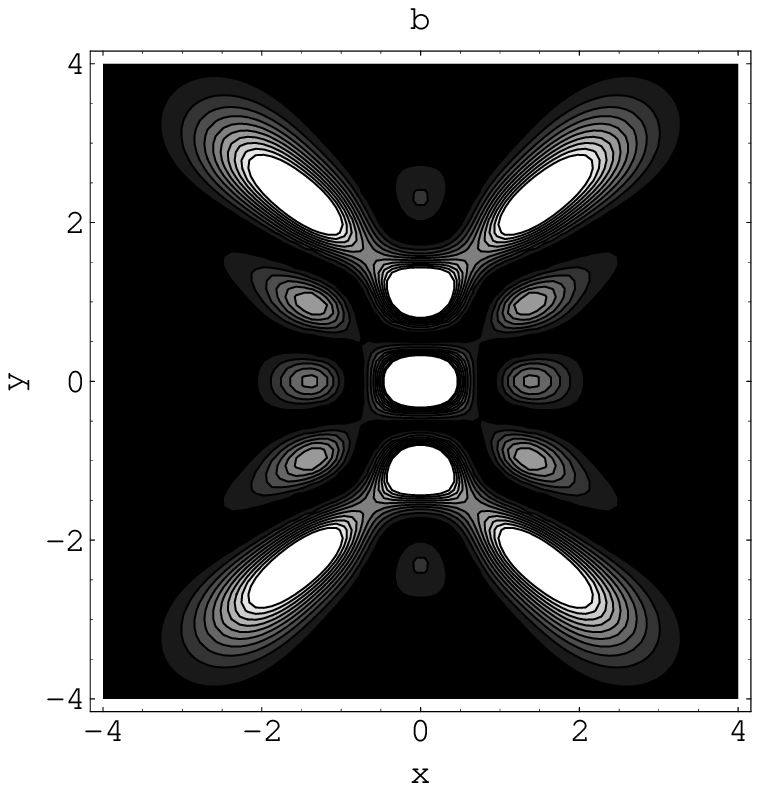} %
\includegraphics[width=7cm]{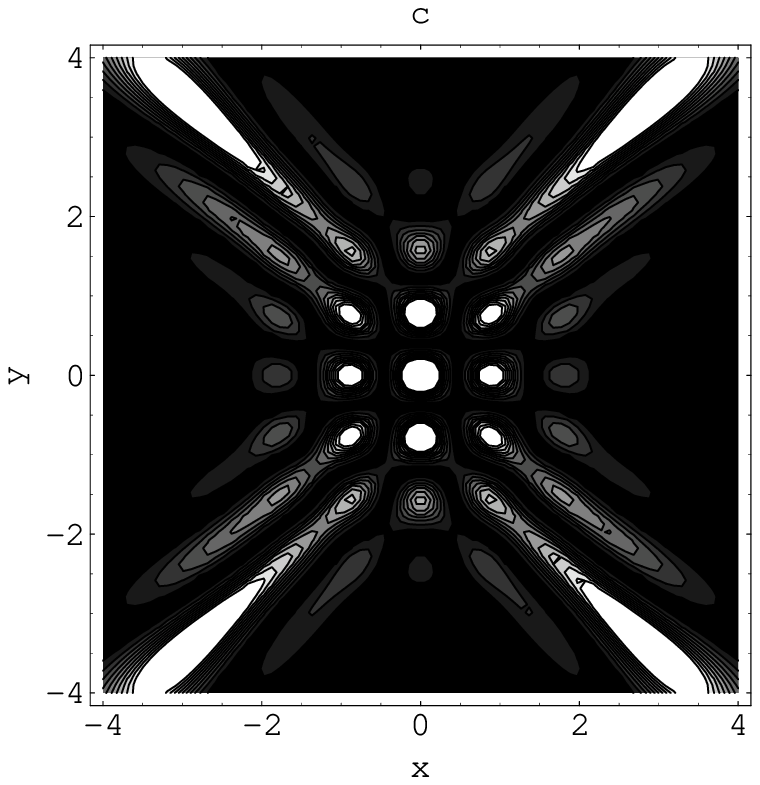}
\end{center}
\par
% Give a unique label
\caption{Contour plot of the quadrature distribution function
$P(x,y)$ for the pair cat state for $q=2, \phi =0$ and for different
values of the parameter $\xi$ where, (a) $\xi=1$, (b)$\xi=3$ and (c)
$\xi=7$.}
\end{figure}

In figure 2, we plot the quadratic distribution for different values
of the number of quant $q$, (say $q=2$, $5$ and $10$). We have shown
that the behavior of the quadratic distribution makes possible, for
sufficient small values of the indicator parameter $q$, to transform
the multi-parts windows to a one circular part. This is a
consequence of the intrinsic properties of the pair cat state.
Nevertheless, perfect purification is not achieved in this case,
compared with the case in which small value of $\xi$ is considered
(see figure 1a and 2a). In figure 1a the window is not a two parts
function, but is very narrow so that the conditional moment is
approximately equal to one circle, this behavior may be obtained if
the parameter $q$ takes smaller values. Note that the quadratic
distribution is dramatically different from the case when we
consider large values of $q$ (see figure 2a and 2b), and indeed here
it is not possible to observe separate parts of the distribution. It
is interesting to show in this figure that the regions of the
quadrature distribution in the $(x,y)$-plane are symmetric with
respect to $y=0$ and with respect to $x=0$ (figures 2b and 2c).

In figure 3, we set the parameter $\phi =0$ and take different
values of $q$ and $\xi .$ In this figure we show the evolution of
$P(x,y)$ and its responses in a type of plot similar to that shown
in figures 1 and 2.  It is shown that the regions of the quadrature
distribution are symmetric with respect to $x=0$ and with respect to
$y=0$ (figures 3b and 3c), but the general feature of this case is
quit complicated compared with the previous cases. This can be
understood from looking to equation (\ref{nphi}), where, when we set
$\phi =0$ this allows the contribution of the second term in
equation (\ref{nphi}).
 Some of the differences are seen in the contour interval if a very small change
 of the controller parameters ($\xi$ and $q$) is considered.
 This can be demonstrated from a contour plot, such as shown in
 figure 3.

\section{Physical system and model Hamiltonian}

In the interest of retaining as much clarity as possible, we first recall
some well-known facts about the system we wish to treat here, which consists
of a single trapped ion interacting with a laser field \cite{gou96}. We
consider the situation where oscillating classical electromagnetic fields
propagating in two-dimensional trap irradiate a single trapped ion. To see
that let us consider the laser frequency is tuned to the frequency
difference between the two modes of the trap. Consequently the Hamiltonian
for such a system is given by \cite{gou96,gou96a,ger98},
\begin{equation}
\hat{H}=\hat{H}_{F}+\hat{H}_{A}+\hat{H}_{in},  \label{h1}
\end{equation}%
where
\begin{eqnarray*}
\hat{H}_{F} &=&\hbar \omega _{1}\hat{a}_{1}^{\dagger }\hat{a}_{1}+\hbar
\omega _{2}\hat{a}_{2}^{\dagger }\hat{a}_{2},\qquad  \\
\hat{H}_{A} &=&\frac{\hbar \omega _{0}}{2}\hat{\sigma}_{z}, \\
\hat{H}_{in} &=&-\wp .E=-\wp .\varepsilon (vt)[\exp [i(k_{1}\hat{x}+k_{2}%
\hat{y}-\omega t))]+H.c.],
\end{eqnarray*}%
with $\wp $ the dipole matrix element of the ion $\wp =\wp ^{0}(\widehat{%
\sigma }_{+}+\widehat{\sigma }_{+}),$ $\varepsilon (vt)$ is a
modulated amplitude of the irradiating laser field. We denote by
$\omega _{i},$ ($i=1,2)$ the fields frequencies, $\omega _{0}$ the
natural frequency of the ion, and $v
$ denotes the ionic velocity. Here the $\sigma ^{\prime }s$ are the usual $%
2\times 2$ Pauli matrices satisfying
\begin{equation}
\lbrack \hat{\sigma}_{z},\hat{\sigma}_{\pm }]=\pm 2\hat{\sigma}_{\pm },[\hat{%
\sigma}_{+},\hat{\sigma}_{-}]=\hat{\sigma}_{z}.
\end{equation}%
We denote by $\hat{a}_{i}$ and $\hat{a}_{j}^{\dagger }$ the Bose operators
for the quantized field mode which obey $[\hat{a}_{i}^{\dagger },\hat{a}%
_{j}]=\delta _{ij},$ where $\delta _{ij}=1$ if $i=j$ and $\mathbf{0}$
otherwise. The operators $\hat{x}$ and $\hat{y}$ are the center of mass
position of the ion which are quantized in the forms $\hat{x}=\Delta x(\hat{a%
}_{1}^{\dagger }+\hat{a}_{1}),$ $\hat{y}=\Delta y(\hat{a}_{2}^{\dagger }+%
\hat{a}_{2})$ and $\Delta x=\sqrt{1/(2v_{x}M)},$ $\Delta
y=\sqrt{1/(2v_{y}M)} $ with $M$ being the mass of the trapped ion.

Making use of the a special form of Baker-Hausdorff theorem
\cite{blo92} the operator $\exp [i\eta (\hat{a}^{\dagger
}+\hat{a})]$ may be written as a product of operators i.e.
\begin{eqnarray}
e^{i\eta (\hat{a}^{\dagger}+\hat{a})}&=&
\exp\left(\frac{\eta^2}{2}[\hat{a}^{\dagger
},\hat{a}]\right)\exp\left(i\eta\hat{a}^{\dagger}\right)\exp\left(i\eta\hat{a}\right)
\nonumber
\\
&=& e^{-\eta ^{2}/2} \sum_{n=0}^{\infty }\frac{\left( i\eta \right) ^{n}\hat{a}%
^{\dagger n}}{n!}\sum_{m=0}^{\infty }\frac{\left( i\eta \right) ^{m}%
\hat{a}^{m}}{m!}.
\end{eqnarray}
Using equation (10), the interaction part of equation (8) can be
written as
\begin{eqnarray}
\hat{H}_{in} &=&-\wp ^{0}.\varepsilon (vt)\eta _{1}\eta _{2}\exp \left[ -%
\frac{1}{2}(\eta _{1}^{2}+\eta _{2}^{2})\right]   \nonumber \\
&&\times \sum_{l=0}^{\infty }\sum_{m=0}^{\infty }\frac{\hat{a}_{1}^{\dagger
l}\hat{a}_{1}^{l}\hat{a}_{2}^{\dagger m}\hat{a}_{2}^{m}}{(l+1)!(m+1)!}(i\eta
_{1})^{l}(i\eta _{2})^{m}  \nonumber \\
&&\times \left( \hat{a}_{1}^{\dagger }\hat{a}_{2}+\hat{a}_{2}^{\dagger }\hat{%
a}_{1}\right) (\hat{\sigma}_{-}+\hat{\sigma}_{+}),  \label{3}
\end{eqnarray}%
with $\eta _{1}=k_{1}\Delta x$ and $\eta _{2}=k_{2}\Delta y.$ When
we take Lamb-Dicke limit and apply the rotating wave approximation
then the effective interaction Hamiltonian (10) takes the two photon
bimodal interaction form. However, provided we adjust the strength
of the electric field to include the time dependent factor
$\varepsilon (vt).$ In this context we may refer to the work given
in Ref. [22]. In this work the authors considered the usual JCM
taking into account the effects of stochastic phase fluctuations in
the atom-field coupling coefficient. This is achieved by modulated
the coupling parameter to include the time factor, in this case it
is observed an interesting feature such as the decoherence effect in
the collapse and revival phenomenon of the Rabi oscillations.

In the Lamb-Dicke regime, $\eta _{j}<<1$, equation (\ref{3}), can be
well
approximated by expanding the exponential terms up to second order in $%
\eta $ ($\eta =\eta _{1}=\eta _{2}),$
\begin{equation}
\hat{H}_{in}=\hbar \lambda (t)\hat{a}_{1}^{\dagger }\hat{a}_{2}\hat{\sigma}%
_{-}+H.c,  \label{eff}
\end{equation}%
where $\lambda (t)=-\varepsilon (vt)\wp ^{0}\eta ^{2}\exp \left(
-\frac{\eta ^{2}}{2}\right) .$
In what follows, we will find the
solution of the equations of motion for
the operators $\hat{a}_{i}(t),$ $\hat{\sigma}_{\pm }(t)$ \ and $\hat{\sigma}%
_{z}(t)$ in the Heisenberg picture. Therefore the equations of motion
related to the Hamiltonian (8) can be written as
\begin{eqnarray}
\frac{d\hat{a}_{1}}{dt} &=&-i\omega _{1}\hat{a}_{1}-i\lambda (t)\hat{a}_{2}%
\hat{\sigma}_{-},\qquad  \nonumber \\
\frac{d\hat{a}_{2}}{dt} &=&-i\omega _{2}\hat{a}_{2}-i\lambda (t)\hat{a}_{1}%
\hat{\sigma}_{+}  \nonumber \\
\frac{d\sigma _{-}}{dt} &=&-i\omega _{0}\sigma _{-}+i\lambda (t)\hat{a}_{1}%
\hat{a}_{2}^{\dagger }\sigma _{z}, \\
\qquad \frac{d\sigma _{z}}{dt} &=&2i\lambda (t)\left( \hat{a}_{1}^{\dagger }%
\hat{a}_{2}\hat{\sigma}_{-}-\hat{a}_{1}\hat{a}_{2}^{\dagger }\hat{\sigma}%
_{+}\right) .  \nonumber
\end{eqnarray}%
After straightforward calculations, one can find an analytic time dependent
solution as%
\begin{eqnarray}
\hat{a}_{1}(t) &=&e^{i\hat{\mu}\alpha (t)}\left( \cos (\gamma _{1}\alpha
(t))-\frac{i\hat{\mu}}{\gamma _{1}}\sin (\gamma _{1}\alpha (t))\right) \hat{a%
}_{1}(0)  \nonumber \\
&&-\frac{i}{\gamma _{1}}e^{-i\omega _{1}t}e^{i\hat{\mu}I(t)}\sin (\gamma
_{1}\alpha (t))\hat{a}_{2}(0)\hat{\sigma}_{-}(0), \\
\hat{a}_{2}(t) &=&e^{-i\omega _{2}t}e^{i\hat{\mu}\alpha (t)}\left( \cos
(\gamma _{2}\alpha (t))-\frac{i\hat{\mu}}{\gamma _{2}}\sin (\gamma
_{2}\alpha (t))\right) \hat{a}_{2}(0)  \nonumber \\
&&-\frac{i}{\gamma _{2}}e^{-i\omega _{2}t}e^{i\hat{\mu}\alpha (t)}\sin
(\gamma _{2}\alpha (t))\hat{a}_{1}(0)\hat{\sigma}_{+}(0), \\
\hat{\sigma}_{-}(t) &=&-\frac{i}{\gamma _{3}}e^{-i\omega _{12}t}e^{i\hat{\mu}%
\alpha (t)}\sin (\gamma _{3}\alpha (t))\hat{a}_{1}(0)\hat{a}_{2}^{\dagger
}(0)  \nonumber \\
&&+e^{-i\omega _{12}t}e^{i\hat{\mu}\alpha (t)}\left( \cos (\gamma _{3}\alpha
(t)) \right.  \nonumber \\
&& \left.+\frac{i\hat{\mu}}{\gamma _{3}}\sin (\gamma _{3}\alpha (t))\right)
\hat{\sigma}_{-}(0), \\
\hat{\sigma}_{z}(t) &=&\cos (2\gamma _{4}\alpha (t))\hat{\sigma}_{z}(0)+%
\frac{i}{\gamma _{4}}\sin (2\gamma _{4}\alpha (t))  \nonumber \\
&& \times\left( \hat{a}_{2}^{\dagger }\hat{a}_{1}\hat{\sigma}_{+}(0)-\hat{a}%
_{1}^{\dagger }\hat{a}_{2}\hat{\sigma}_{-}(0)\right),
\end{eqnarray}%
where $\omega _{12}=\omega _{1}-\omega _{2},$
\begin{eqnarray*}
\gamma _{1} &=&(\hat{l}+\frac{1}{2})(\hat{m}+\frac{3}{2}),{\ \ }\gamma _{2}=(%
\hat{l}+\frac{3}{2})(\hat{m}+\frac{1}{2}), \\
\gamma _{3} &=&(\hat{l}-\frac{1}{2})(\hat{m}+\frac{3}{2}),{\ \ }\gamma _{4}=(%
\hat{l}+\frac{1}{2})(\hat{m}+\frac{1}{2}), \\
\alpha (t) &=&\int_{0}^{t}\lambda (t^{^{\prime }})dt^{^{\prime }},{\ \ }\hat{%
\mu}=(\hat{a}_{1}^{\dagger }\hat{a}_{2}\hat{\sigma}_{-}+\hat{a}_{2}^{\dagger
}\hat{a}_{1}\hat{\sigma}_{+}).
\end{eqnarray*}%
Having obtained the dynamical operators, we are therefore in a position to
discuss some statistical properties of the system.

\section{Atomic inversion}

The entanglement of motional degrees of freedom of the center of mass of the
ion with it's internal degrees of freedom manifests itself in the well known
collapse and revival of population inversion \cite{sha02}. We shall start
with the atomic inversion from which we can discuss the collapse and revival
phenomenon. The (internal level) ionic dynamics depend on the distributions
of initial excitations of both the field and the center-of- mass vibrational
motion, given by $\langle n|\rho _{f}(0)|n\rangle =\rho _{nn}^{f}(0)$ and $%
\langle m|\rho _{v}(0)|m\rangle =\rho _{mm}^{v}(0)$, respectively.
For instance, the atomic population inversion may be written as
\begin{equation}
\langle \sigma _{z}(t)\rangle =Tr\left( \sigma _{z}(0)\rho
(0)\right) .
\end{equation}%
that is the difference between the probability of finding the system in the
ground state and the probability of finding the system in the excited state.

To analyze the effects resulting from variation in the parameter $\lambda (t)
$ on the atomic inversion we have plotted in figure 4 the function $\langle
\sigma _{z}(t)\rangle $ against the scaled time $\lambda t,$ for different
values $\lambda (t),$ where $\lambda (t)=\lambda $ for figure (4a) and $%
\lambda (t)=\lambda \sinh (\varpi t)$ for figure (4b), where $\varpi
$ is a positive integer. In this figure let us take the parameters
$\xi =10,q=1$ and $\phi=\pi/2$, keeping in mind we are dealing with
the problem at exact resonances. In figure (4a) we find that the
value of the atomic inversion decreases from the maximum to its
minimum, (approximately $-0.75)$ then it collapses, but then it
starts to fluctuate around zero for a short period of time. Thus the
Rabi frequencies become commensurate so that the
atomic inversion has an exact periodic evolution with a scaled $\lambda $%
-dependent period. In this case there is no oscillations in the
collapse region, as shown in figure 4a. A different pattern for the
oscillations occurs if the number difference is slightly changed.
\begin{figure}[tbp]
%\vspace*{3cm}
\includegraphics{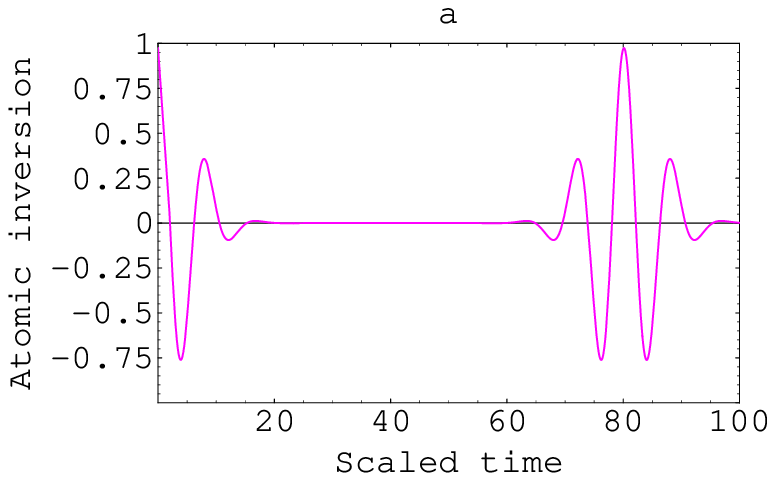} \includegraphics{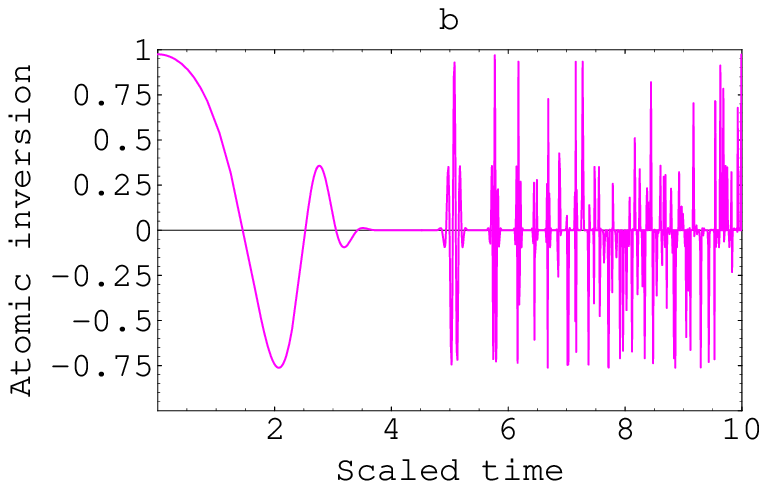} % Give a unique label
\caption{The atomic inversion as a function of the scaled time $%
\lambda t$. Parameters: $\xi =10,q=1,\phi =0,$ where (a) $%
\lambda(t)=\lambda t$ and (b) $\lambda(t)=\lambda\sinh(\varpi t)$}
\label{fig:2}
\end{figure}
Strictly speaking, when we consider the time dependent modulating
function $\lambda (t)=\lambda \sinh (\varpi t),$ the function $\sinh
(\varpi t)$ oscillates (however their period of oscillation is not
constant) then the appearance of the collapses and revivals in
almost a periodic way has been washed out in this case. On the other
hand and in the presence of the time-dependent modulating function
we find the behavior of the atomic inversion function is drastically
changed where we can see irregular collapses and revivals occur,
(see figure 4b). The short time revivals are strongly suppressed, as
we see in figure 4b. We may compare the atomic inversion in both
figure 3a and 3b, in the former case the collapse time is long
enough to still allow revivals, while in the latter a shorter
collapse time with the same amplitudes of the revivals.

It is interesting to explore to what conditions can be considered to
generate a long living entanglement. This will be seen in the next
section.

\section{Entanglement}

It is well known that entanglement between several particles is the
most important feature of many quantum communication and computation
protocols. In order to get a feel for how the entanglement of
ion-field interaction affected by the presence of pair cat states we
began  by studying the aforementioned system that is analytically
solvable and start from a factored initial state of both the ion and
field. We consider the notion of entanglement entropy, that is,
considering a quantum system in a pure state with the density matrix
of the remaining space as $\rho _{A}=Tr_{F}\left( \rho \right) $.
The density matrix contains all information of any system in a mixed
or pure state, and computing vital physical information on any such
system is mostly determined by the eigenvalues of the density
matrix. The partial entropy of entanglement for a bipartite pure
state is defined as Von Neumann entropy of the reduced state,
\begin{equation}
S_{A}=-Tr\left\{ \rho _{A}\log _{2}\rho _{A}\right\} .
\end{equation}%
This procedure is well established, and works well for all cases
where the initial state of the system is in a pure state. Most work
has been focused on this entanglement in time-independent modulating
function, but we will focus on a time-dependent case.

On the other hand, the quantum entropy of the field $S_{f}(t)$ can be
expressed in terms of the eigenvalues $\lambda _{F}^{\pm }(t)$ of the
reduced field density operator as follows
\begin{equation}
S_{f}(t)=-[\lambda _{f}^{+}(t)\ln \lambda _{f}^{+}(t)+\lambda _{f}^{-}(t)\ln
\lambda _{f}^{-}(t)].  \label{28}
\end{equation}%
In the case of a disentangled pure joint state $S_{f}(t)$ is zero, and for
maximally entangled states it gives $\ln 2$ \cite{pho88}. Due to the higher
dimensionality of the problem we cannot obtain a simple analytical
expression for equation (\ref{28}), therefore the numerical approach becomes
indispensable.

\begin{figure}[tbp]
%\vspace*{3cm}
\includegraphics{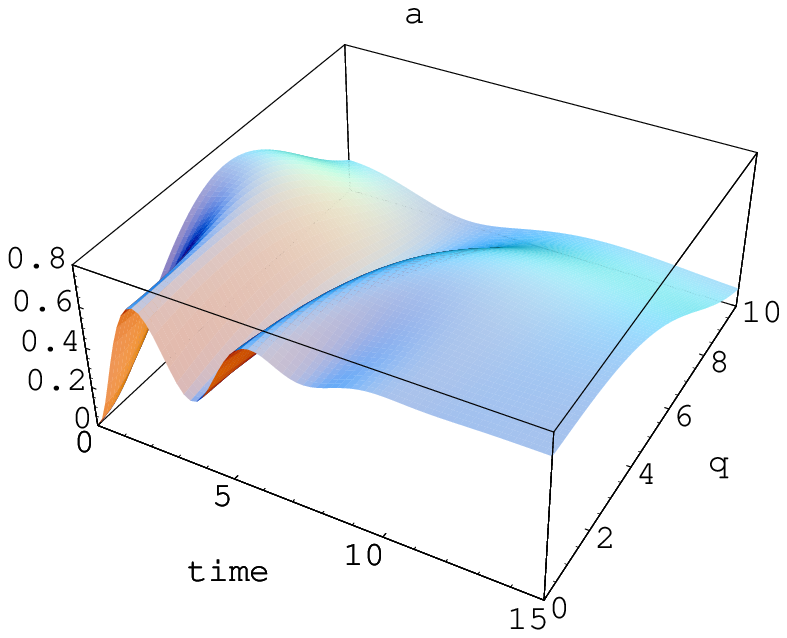}
\includegraphics{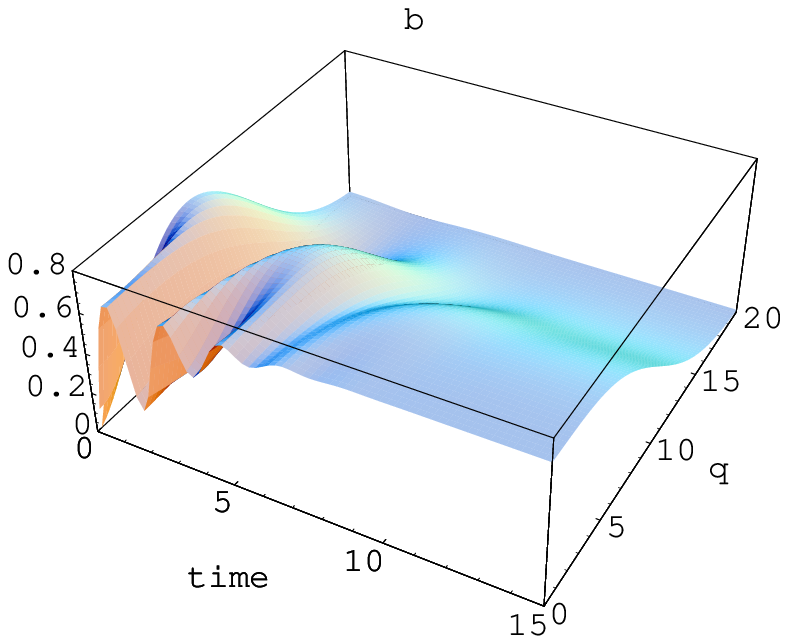}
\caption{The quantum field entropy $S_{f}(t)$ \ as a function of the scaled
time $\lambda t$. The parameters are $q=1,$ $\eta =0.2,$ $%
\lambda (t)=\lambda $ and different values of $\xi $ where (a) $\xi
=10$ and (b) $\xi =20.$} \label{fig:2}
\end{figure}

Now let us discuss the behavior of the field entropy resultant of
the existence of the time dependent modulating function
$\lambda(t)$. Figure 5 is a plot of
quantum field entropy $S_{f}(t)$ as a function of scaled time parameter $%
\lambda t$ \ for $\eta =0.2,$ $\lambda (t)=\lambda $ and considering
different values of $\xi $ where $\xi =10$ for figure 5a and $\xi
=20$ for figure 5b. Our numerical analysis was performed using other
parameters from recent experiments \cite{mon96,mee96}, where a
trapped $^9$Be$^+$ ion was laser-cooled to the zero-point energy and
then prepared in a superposition of spatially separated coherent
harmonic oscillator states. In \cite{mun02} there have been recent
demonstrations of the coupling between cavities and single-ion
traps, where, a single calcium ion is stored in a spherical trap and
placed in the center of a confocal resonator.

It is interesting to note that, the maximum entanglement is
decreases as the parameter $q$ increased (see figure 5a). As we see
from figure 5a, as time goes on we note a growth in $S_{f}(t)$,
followed by a sudden decrease, almost down to zero at large values
of $q$. When the time increases further we see that the entanglement
has a long surviving, if we keep smaller values of $q$. The constat
value of $S_{f}(t)$ is interpreted as a result of quantum
entanglement between ion and fields, surviving for a long time
interaction. However, an interesting situation may arise when the
parameter $\xi$ takes large values (say $20$ in figure 5b). In this
case, for some small fixed values of $q$, the entanglement surviving
starts at earlier time. However, the overall features of the process
are similar in both small and large values of $\xi$. An important
difference occurs only at earlier interaction time, where the number
of oscillations is increased with increasing $\xi$. Also, the
amplitudes of these oscillations are bigger compared to the case in
which $\xi=10$ (see figures 5a and 5b). As seen in the entropy plot
in figure 5b, the pure state is basically obtained for large values
of the parameter $q$ (because the natural scale of $q$ is between
$0$ and $\xi$, in figure 5b, we expand the scale of $q$ to take
values between $0$ and $20$).

On the contrary, as can be seen from figure 5, if the fields are
initially prepared in a pair cat state and the modulating function
is taken to be time-dependent $\lambda (t)=\lambda \sinh (\varpi
t)$, then the entanglement parameter is significantly larger than
zero for any time $t>0$, but with some oscillations from time to
time. Also, the number of oscillations is increased as the parameter
$q$ decreased associated with low entanglement reaches zero for
smaller values of the parameter $\xi $ (see figure 6).

\begin{figure}[tbp]
%%\vspace*{3cm}
\includegraphics{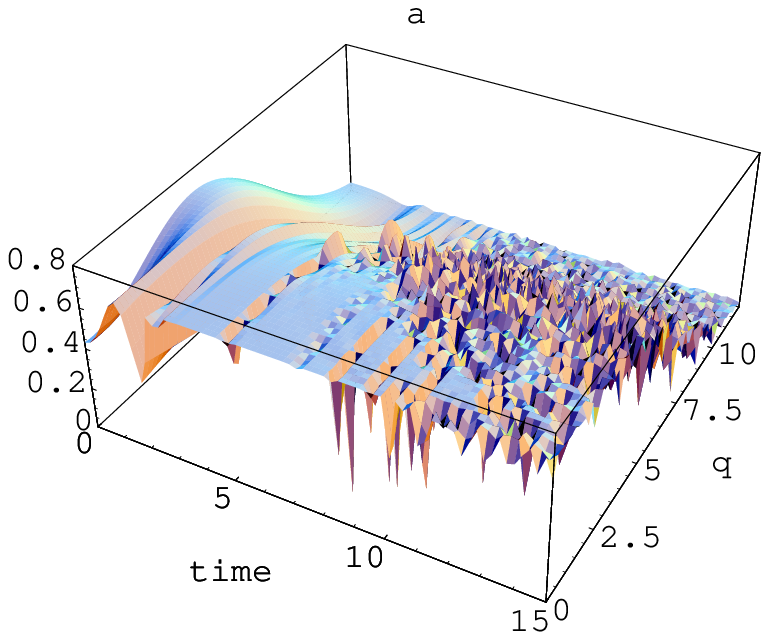} \includegraphics{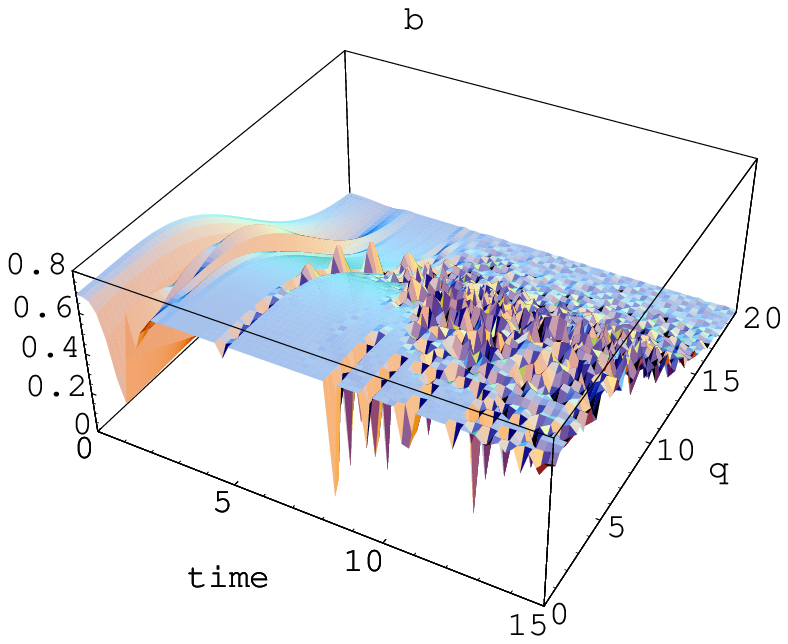} % Give a unique label
\caption{The same as figure 5 but $\lambda (t)= \lambda %
\sinh (\varpi t)$.} \label{fig:2}
\end{figure}
Another interesting aspect is that the time-dependent modulating
function $\lambda(t)$ has an important affect on the entanglement,
where, the total ion-field state can not have its purity diminished,
which means that as the field becomes more pure the ionic state must
be closer to a mixed state. Also, in this case, it seems remarkable
that entanglement survive for long time but some oscillations occur
here. More surprisingly still, survive is long-lived for the
time-independent case. We should pointed out that if we simply use a
time-independent modulating function, we would expect entanglement
to survive.

Practically speaking, as can be deduced from figure 5a, the
oscillations in degree of entanglement between the ion and the field
quickly damp out with an increase in $q$. The subsystem will not
disentangle from each other until the steady state is reached as
long as the interaction is present. This means that neither the ion
nor the field will return back to a pure state except at the
beginning of the interaction and $t\rightarrow \infty ,$ contrast to
the $q=0$ case. In order to gain insight into the general behavior
we have considered different values of $\xi $ in figure 6b$.$ The
direct comparison of figure 6a and figure 6a, the amplitude of the
oscillations in the degree of entanglement is increased with
increasing the parameter $\xi $.

The remaining task is to identify and compare the results presented
above for the entanglement degree with another accepted entanglement
measure such as the linear entropy. The question of the ordering of
entanglement measures was raised in reference \cite{eis99}. It was
proved that all good asymptotic entanglement measures are either
identical or fail to uniformly give consistent orderings of density
matrices \cite{vir00}. The best understood case, not surprisingly,
is the simplest. In order to better characterize the subsystems, we
may refer here to another measure of the entanglement of a reduced
density matrix which is the product state identification
\[
S_{L}=1-Tr\rho _{A}^{n},{\ \ \ \ \ \ \ \ }n\geq 2,
\]%
which is zero for a product state, and unity for a maximally entangled
state. This measure is equivalent to the R\u{e}nyi entropy \cite{skr05} and
is not well suited for much more than to single out a pure state, as with
increasing $n$ any entangled state will converge to zero in this measure.

In figure 7 we plotted the entanglement according to a high order
linear entropy and compared with the von Neumann entropy. One,
possibly not very surprising, principal observation is that the
numerical calculations corresponding to the same parameters, which
have been considered in figures 4 and 5, gives nearly the same
behavior (see figure 7). This means that both the entanglement due
to the quantum field entropy $S_f$ and high order linear entropy
$S_L$ measures are qualitatively the same. The important consequence
of this observation is that the use of quantum entropy and linear
entropy as measures of entanglement are equivalent. We must stress,
however, that no single measure alone is enough to quantify the
entanglement in a multi-partite systems. Finally, we may say that,
it is possible to obtain a long living entanglement using the
time-independent modulating function in pair cat state. This result
is quit surprising, since the previous studies \cite{pho88} in the
entanglement for the time-independent interaction for the initial
coherent state does not contain this interesting feature. Which
means that the pair cat state as an initial state of the field plays
very important roles in the quantum information process.

\begin{figure}[tbp]
%%\vspace*{3cm}
\includegraphics[width=9cm]{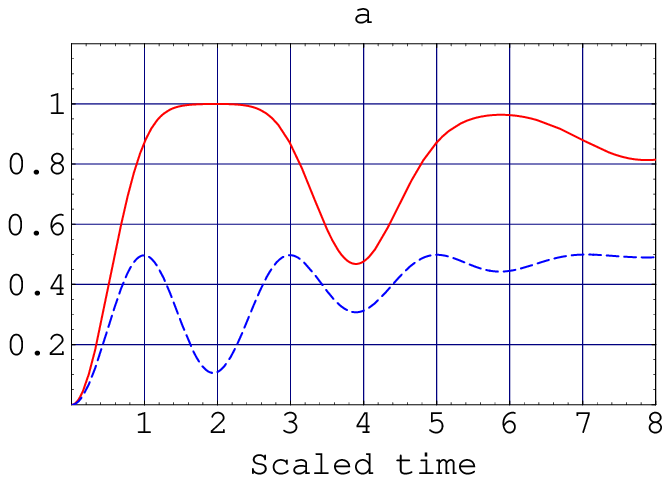} \includegraphics[width=9cm]{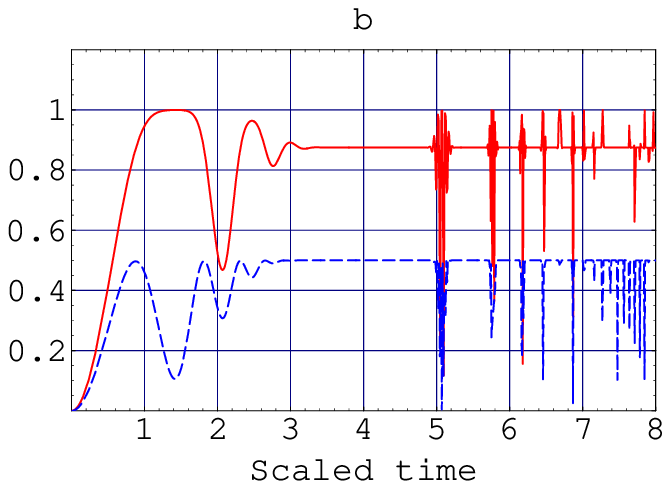}
% Give a unique label
\caption{Time evolution of entanglement degree $S_L$ as a function
of the scaled time $\lambda t$. The parameters are the same as in
figure 3, where $n=2$ (the solid curve) and $n=3$ (the dotted
curve). } \label{fig:2}
\end{figure}

A proposal for the generation of motional pair cat states in a
two-dimensional anisotropic trap has been introduced in Ref. \cite{zhe01}.
In order to generate the pair cat states for the motion in a two-dimensional
anisotropic trap we require two laser beams and the frequencies of the laser
beams are chosen as $\omega _{1}=\omega _{0}-2\nu _{x}-2\nu _{y},$ and $%
\omega _{2}=\omega _{0}$. The generation method requires a sequence
of many measurements of the internal electronic state at different
time instants appropriately chosen \cite{man05}. Apart from
detecting quantum entanglement in these states, might be a useful
way to fully characterize simple quantum logic gates \cite{poy97}.

\section{Summary}

The main results of this work are summarized as follows: After using
the Heisenberg approach to the quantum operator, we have
investigated effects produced by the ion-field interaction. We have
shown that the input pair cat states exhibit strong nonclassical
features due to the correlation between the two modes. The
phenomenon of quantum revivals in the time-dependent ion-field
interaction and the atomic inversion properties as well as the
entanglement have been analyzed. Also, it has been shown that the
atomic inversion as a function of the scaled time may display
different structures of beats depending on the initial state of the
field and the vibrational motion as well as on the time-dependent
interaction. Effects such as suppression or attenuation of the Rabi
oscillations and long time scale revivals as well as a periodic
dynamics have been observed.

Using pair cat states as an initial state of the fields, our
analysis shows that the fluctuations of the entanglement between
maximum and minimum values are irregular for both the
time-independent and time-dependent cases at a short period of the
interaction time. Our results reveal a new feature in the
entanglement as a function of the scaled time, that is a long living
entanglement. This feature depends on the parameters $\xi$ and $q$.
These interesting features may pave the way to quantum information
applications such as deterministic all-optical quantum computation.
The results can be relevant for basic studies on entanglement as
well as for applications in the preparation and manipulation of
quantum states for quantum information purposes. As far as
scalability is concerned, one could think in having several trapped
ions in a high-Q cavity, exchanging information via the cavity
photons, or other more sophisticated schemes involving trapped ions,
phonons and photons.

It is to be remarked that, the entropy of entanglement can be
determined indirectly by experiment using the quantum state
tomography \cite{nie00}. This fact comes from the definition of the
entropy of entanglement which is associated with the density matrix
of quantum state. We believe that the pair cat states may be a
useful tool to be considered in obtaining a long living entanglement
of systems of indistinguishable particles. While our analysis was
carried out with respect to the example of a single trapped ion
beyond Lamb-Dicke regime, we emphasize that the results here
reported apply equally well to any bipartite system. We hope to
report on such issues in a forthcoming paper.

\textbf{ACKNOWLEDGMENTS}

The author would like to thank the referees for their objective
comments that improved the text in many points. Also, it is a
pleasure to thank G. Gour, A.-S. F. Obada, S. S. Hassan, and M. S.
Abdalla for inspiring discussions.

\end{document}